\RequirePackage{fix-cm}
\documentclass[smallextended]{svjour3}       
\smartqed  
\usepackage{graphicx}
\graphicspath{{.}{final_figs/}}
\usepackage[english]{babel}
\usepackage{amssymb}
\usepackage{amsbsy}
\usepackage{amsmath}
\usepackage{bm}
\usepackage{color}
\usepackage{dcolumn}
\usepackage{tabularx}
\usepackage{tikz}
\usepackage{bbm}

\usepackage{xspace}
\newcolumntype{d}[1]{D{.}{.}{#1} }
\usepackage{subcaption}
\usepackage{algpseudocode}

\usepackage{times}
\usepackage{mathptm}
\usepackage{algorithm}
\usepackage{algorithmicx}
\usepackage{algpseudocode}
\usepackage{breakcites}

\usepackage{siunitx}
\usepackage{amsfonts}

\usepackage{natbib} 

\newif \ifnotesw \noteswtrue

\journalname{Mathematical Geosciences}

\begin{document}

\title{Multilevel Graph Partitioning for Three-Dimensional Discrete Fracture Network Flow Simulations}
\titlerunning{Graph Partitioning for 3D DFN  Simulations}        

\author{Hayato Ushijima-Mwesigwa,  Jeffrey D. Hyman, Aric Hagberg, Ilya Safro,  Satish Karra, Carl W. Gable,  Matthew R. Sweeney, and Gowri Srinivasan}

\authorrunning{H. Ushijima-Mwesigwa et al.}

\institute{Hayato Ushijima-Mwesigwa \at Fujitsu Laboratories of America, Inc. Sunnyvale, CA 94085. Work done while at Los Alamos National Laboratory, Los Alamos New Mexico, USA\\
\email{hayato@fujitsu.com}
\and
J. D. Hyman \at
Computational Earth Science (EES-16), Earth and Environmental Sciences Division, Los Alamos National Laboratory, Los Alamos, New Mexico, USA 87505\\
\email{jhyman@lanl.gov}  
\and
      Aric Hagberg \at Computer, Computational, and Statistical Sciences Division, Los Alamos National Laboratory, Los Alamos New Mexico, USA \\
    \email{hagberg@lanl.gov}
\and
Ilya Safro \at Computer and Information Sciences, University of Delaware,  Newark, DE, USA \\
\email{isafro@udel.edu}
\and 
C. W. Gable \at
Computational Earth Science (EES-16), Earth and Environmental Sciences Division, Los Alamos National Laboratory, Los Alamos, New Mexico, USA 87505\\
\email{gable@lanl.gov}
\and
Matthew R. Sweeney\at Computational Earth Science (EES-16), Earth and Environmental Sciences Division, Los Alamos National Laboratory, Los Alamos New Mexico, USA \\
\email{msweeney2796@lanl.gov}
\and
Gowri Srinivasan \at Verification and Analysis (XCP-8),  X Computational Physics, Los Alamos National Laboratory, Los Alamos New Mexico, USA \\
\email{gowri@lanl.gov}
}
\date{~}

\maketitle

 \begin{abstract}
We present a topology-based method for mesh-partitioning in three-dimensional discrete fracture network (DFN) simulations that takes advantage of the intrinsic multi-level nature of a DFN.  DFN models are used to simulate flow and transport through low-permeability fractured media in the subsurface by explicitly representing fractures as discrete entities. The governing equations for flow and transport are numerically integrated on computational meshes generated on the interconnected fracture networks.  Modern high-fidelity DFN simulations require high-performance computing on multiple processors where performance and scalability depends partially on obtaining a high-quality partition of the mesh to balance work-loads and minimize communication across all processors. 

The discrete structure of a DFN naturally lends itself to various graph representations, which can be thought of as coarse-scale representations of the computational mesh.  Using this concept, we develop two applications of the multilevel graph partitioning algorithm to partition the mesh of a DFN. 
In the first, we project a partition of the graph based on the DFN topology onto the mesh of the DFN and in the second, this DFN-based projection is used as the initial condition for further partitioning refinement of the mesh. 
We compare the performance of these methods with standard multi-level graph partitioning using graph-based metrics (cut, imbalance, partitioning time), computational-based metrics (FLOPS, iterations, solver time), and total run time.  
The DFN-based and the mesh-based partitioning methods are comparable in terms of the graph-based metrics, but the time required to obtain the partition is several orders of magnitude faster using the DFN-based partitions. 
The computation-based metrics show comparable performance between both methods so, in combination, the DFN-based partitions are several orders of magnitude faster than the mesh-based partition. 
Moreover, the method which uses the DFN-partition solution as the initial condition of the mesh partition provided cut and imbalance values that were close to the mesh-based partition but in a fraction of the time. 
In turn, this hybrid method outperformed both of the other methods in terms of the total run time. 


\end{abstract}

\section{Introduction}

Discrete Fracture Network (DFN) models are a modeling framework to represent fractured systems most commonly applied to resolve flow and transport through low-permeability subsurface fractured rock. 
DFN models differ from conventional continuum models by  explicitly representing fractures and the networks they form.  
This  allows DFNs to represent a wider range of transport phenomena and makes them a preferred choice when linking network attributes to flow properties~\citep{hadgu2017comparative,hyman2020flow,hyman2018dispersion}.
DFNs are utilized for characterizing fluid flow and solute transport through low permeability fractured media which is critical for a variety of subsurface applications including the environmental restoration of contaminated fractured media~\citep{national1996rock,neuman2005trends,vanderkwaak1996dissolution}, aquifer storage and management~\citep{kueper1991behavior}, hydrocarbon extraction~\citep{hyman2016understanding,karra2015effect,middleton2015shale}, longterm storage of spent civilian nuclear fuel~\citep{frampton2010inference,hadgu2017comparative,joyce2014multiscale}, and CO$_2$ sequestration~\citep{hyman2020characterizing,jenkins2015state}.

The choice to explicitly represent fractures in DFN results in a significantly higher computational cost than stochastic continuum~\citep{neuman2005trends}, dual porosity/permeability \citep{lichtner2014modeling}, and generalized upscaled models~\citep{sweeney2019upscaled} where upscaled effective properties are used to account for fracture properties, cf Berre et al.~\citep{berre2018flow} for a discussion of the benefits and drawbacks of each method. 
Once a network is constructed, the individual fractures are meshed for computation and the governing equations for flow and transport are numerically integrated on the computational mesh.  
Due to the inherent uncertainty in characterizing the subsurface, multiple realizations of each DFN are required, which further increases the computational demand~\citep{osthus2020probabilistic}.  
The number of mesh cells required for a DFN depends on the number of fractures, the density of the network, and the range of length scales being resolved. 
Even modest sized DFNs containing $\mathcal{O}(10^2)$ fractures can have a mesh that contains several million nodes. 
Because of limited computational resources, the first DFN models either represented networks as a set of connected pipes~\citep{cacas1990modeling,dershowitz1999derivation} or were two-dimensional representations~\citep{de2004influence}.   
Despite the large scale of DFN computations, high performance computing (HPC) enables flow and transport simulations in large three-dimensional DFNs~\citep{berrone2013pde,berrone2015parallel,erhel2009flow,hyman2015dfnworks,mustapha2007new,pichot2010mixed,pichot2012generalized}.

These high-fidelity DFN simulations require HPC on multiple processors and the performance and scalability of these simulations necessitate a high-quality partitioning such that the computations are well-balanced across all processes with minimal communication between processors.
Common partitioning methods \citep{bulucc2016recent} are based on either a global  method, such as  spectral partitioning and max-flow, or iterative local improvement heuristic algorithms, such as  Kernighan-Lin \citep{KL} or Fiduccia-Matthesyses \citep{fiduccia1988linear}.   
However, multilevel graph partitioning, which introduces a framework to make global decisions in conjunction with local improvements, is one of the most successful heuristics in practice for partitioning large graphs \citep{chevalier2009comparison,karypis1998fast,ron2011relaxation,safro2015advanced,sanders2011engineering,walshaw2000mesh}.   

The basic idea behind multilevel graph partitioning is that a graph is successively coarsened, creating a hierarchy of smaller graphs until an initial (coarsest) partition can be computed efficiently.
The initial partition is projected back to the next finer level, where local improvements are made. 
Once at a local optima, the improved partition is projected to the next finer level where further local improvement are made. The process continues until a partition is projected and refined back to the original graph. 
Multilevel graph partitioning methods are popular because they exhibit excellent trade-off between fast computational time and high-quality solutions compared to other techniques. 
However, some applications (for example those involving dynamic graphs) require graphs to be repartitioned, and thus require much faster techniques. 
Thus, depending on the application, even multilevel graph partitioning can take a significant amount of time. 

In order to increase the speed of high-resolution DFN simulations, we propose a graph partitioning approach for the DFN mesh that combines the topological structure of the DFN with multilevel graph partitioning.  
The discrete structure of a DFN naturally lends itself to various graph representations, for example, vertices in graph can correspond to fractures in the DFN and edges in the graph to  fracture intersections.
These graph-representations of a DFN can also be thought of as coarse-scale representations of the computational mesh which is the conceptual model that we use here to develop applications of the multilevel graph partitioning algorithm for mesh partitioning.
By using this partitioning of the DFN, rather than the mesh, we seek to accelerate the initial setup of HPC computations.
We consider two methods where the first coarse level in the multilevel graph partitioning is a weighted graph based on the topology of the DFN that accounts for the number of mesh nodes on each fracture. 
In one method, we directly project the solution returned by the multilevel graph partitioning algorithm onto the DFN mesh, while in the second, this solution is used at the initial condition for the mesh partitioning. 
In Berrone et al.~\citep{berrone2015parallel}, the topology of the fracture network was used as a coarse partitioning for parallel computation.
However, in that previous work, the task of partitioning was not considered a major issue and a robust solution in this direction was not explored in depth.

We compare the relative cost of the proposed method with partitioning the full mesh and find that the total run time is reduced by several orders of magnitude using the proposed methods. 
Partitioning the graph-representation of the DFN and projecting the solution onto the mesh is computationally cheaper than partitioning the DFN mesh itself since there are orders of magnitude fewer nodes and edges to consider in a graph based on the DFN topology.
The method is also sensitive to the mesh resolution on each fracture, such as  it accounts for the number of mesh nodes on each fracture. 
The performance of the method compared to partitioning the mesh is measured in terms of graph-based metrics (cut, imbalance, partitioning time), computational-based metrics (FLOPS, iterations, solver time), and total run time. 
In terms of graph-based metrics, the results obtained using the DFN-based partitions are comparable  to those obtained using the mesh-based partition, yet the DFN-based partition is several orders of magnitude faster.
However, the method that further refines the mesh using the DFN partition as the initial condition provides cut and imbalance values that are close to the mesh-based partition in a fraction of the time. 
This hybrid method outperformed both of the other methods in terms of the balance between run time and mesh partition quality.
This additional refinement should be applied as its additional cost is negligible compared to other aspects of the simulation, but the improved performance of the solver is substantial. 
The results presented here indicate that using the proposed hybrid method reduces the overall computation time of a single DFN realization simulation, and consequently allows computation of more realizations for uncertainty quantification for a fixed computational budget.

\section{Discrete Fracture Networks}

Fluid flow and the associated transport of solutes in low-permeability fractured media, such as shale and crystalline rock, is primarily confined to the fractures surrounding by a solid rock matrix~\citep{national1996rock}.
Within the fractured media, the geometry of individual fractures, such as  their size and aperture, along with the network structure, such as  connectivity and density, rather than matrix properties, such as  matrix porosity or pore-size distributions, determine the structure of the fluid velocity field~\citep{davy2013model,dreuzy2012influence,hyman2018dispersion,hyman2019linking}.
There are several computational methods used to model flow and the associated transport of chemical species.
There are stochastic continuum~\citep{neuman1988use,neuman2005trends,tsang1996tracer} methods, which use effective upscaled properties to capture the effects of the fractures. 
There are dual-porosity / dual-permeability~\citep{gerke1993dual,zimmerman1993numerical,lichtner2014modeling} models, which partition the domain into multiple domains connected via transfer functions.
There are also discrete fracture network models (DFN)~\citep{cacas1990modeling,long1982porous,maillot2016connectivity,makedonska2016evaluating,nordqvist1992variable}, which is the model we focus on in this study. 
Within a DFN model, every individual fracture is represented as a planar $N-1$ dimensional object embedded within an $N$ dimensional space. 
For example, in two-dimensions, fractures are lines, and in three-dimensions, fractures are planes.

However, the generation of explicit fracture representations results in some complications. 
Foremost, conditioning the generation of a DFN model requires detailed hydraulic and structural information, which, due to the size of the domains of interest, up to multiple kilometers, along with the monetary cost of sufficient sampling of relevant quantities in the subsurface~\citep{bonnet2001scaling,national1996rock,zimmerman1993numerical}, requires that DFN models are stochastically constructed.
Using information obtained from a site characterization, attributes of each fracture, such as  size and orientation, are sampled from calibrated distributions~\citep{joyce2014multiscale}.
The fractures interconnect to form a network.
Every fracture in the network is meshed for computation, and the governing equations for flow and transport are numerically integrated to simulate a particular physical process of interest.
 
\begin{figure}[htb!] \centering{
 \includegraphics[width=0.75\textwidth]{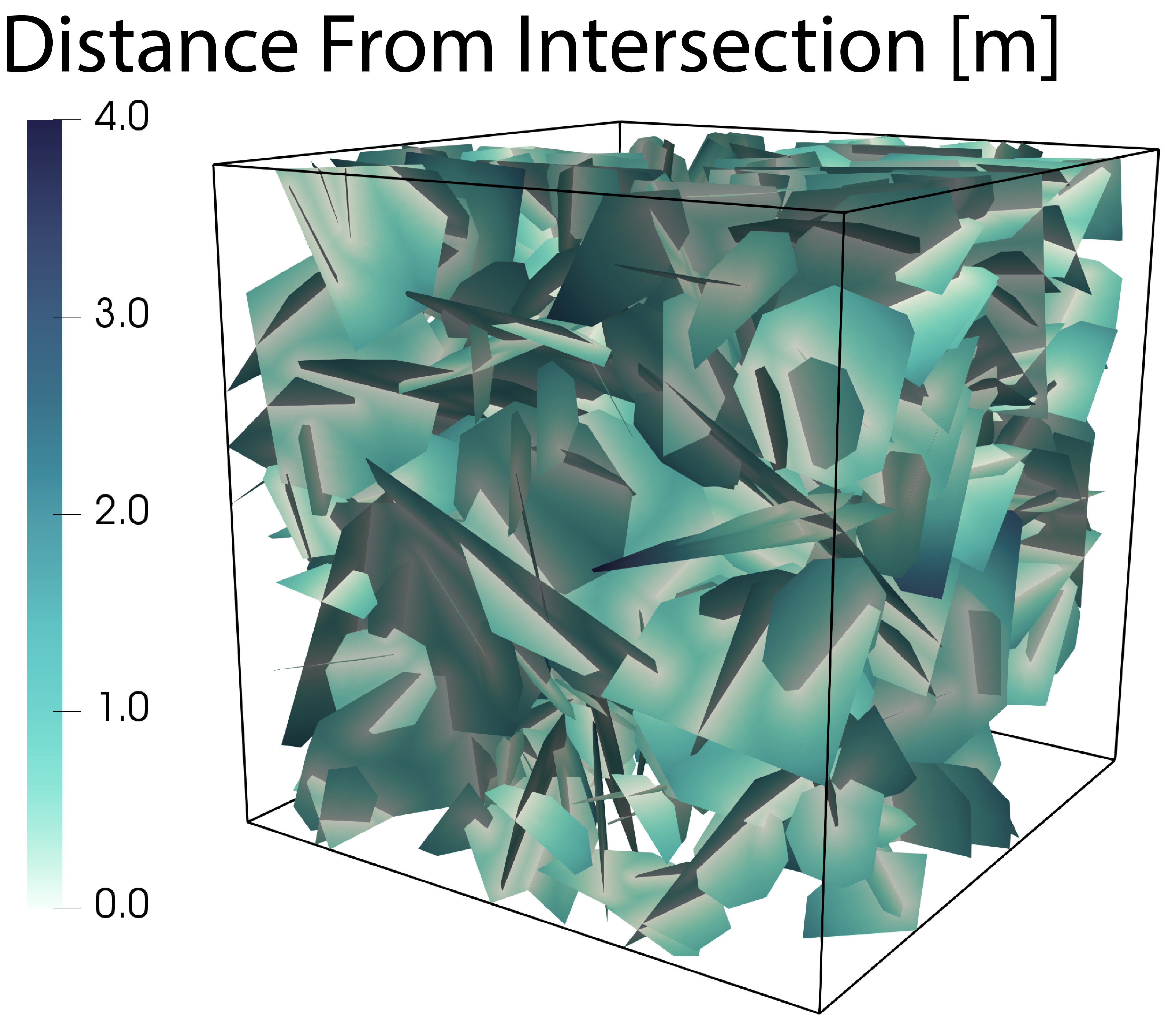} \\
}
\caption{\label{fig:dfn} A Discrete Fracture Network (DFN) composed of 424 fractures in a 15 meter cube where colors correspond to the distance on the fracture plane to the nearest fracture intersection. The regions colored white are close to fracture intersections and  darker colors indicate larger distances. The variability in colors on a single plane highlights the range of length scales that exist on a single fracture and throughout the network. To properly simulate relevant physical phenomenon, the mesh representation of the network must be fine enough to resolve all of these length scales}
\end{figure}

The stochastic generation of a DFN is a major obstacle in the creation of a high-quality computational mesh representation of each network. 
In practice, the planes representing each fracture are randomly included into the domain and can create arbitrarily small length features that render the automated meshing of the fracture planes incredibly challenging.
Figure~\ref{fig:dfn} shows a DFN composed of 424 fractures in a 15 meter cube, which are represented as circular polygons, to demonstrate the range of length scales that exists in a DFN.
There are 494 intersections between fractures in the DFN.
Colors on the fractures correspond to the distance on the fracture to the nearest line of intersection and highlight the range of length scales that exists on a fracture plane and throughout the network.
Mesh edges close to these features must be smaller that the smallest local length scale if the geometry and physics are to be properly resolved.
This requirement is computationally infeasible for arbitrarily small length scales within large domains.
There have been a number of methodologies to address this issue by modifying the mesh to remove small features~\citep{mustapha2011efficient,mustapha2007new} or coupling flow between non-conforming meshes using discretization schemes~\citep{berrone2013pde,erhel2009flow,pichot2010mixed,pichot2012generalized}. 


The Feature Rejection Algorithm for Meshing ({\sc fram}) introduced by Hyman et al.~\citep{hyman2014conforming} is one method designed to address the mesh generation issues above by constraining the generation of the network so that the smallest features are larger than a user defined minimum length scale.
Therefore {\sc fram} allows for detailed control of the mesh resolution on each fracture because pathological cases that degrade mesh quality do not exist. 
A result of using {\sc fram} is a conforming Delaunay triangulation on each fracture, where conforming means that all lines of intersection form a set of connected edges in the mesh, can be automatically generated using the method detailed in Murphy et al.~\citep{murphy2001point}.
Depending upon the physical process to be simulated, the mesh can either have variable or uniform resolution. 
Points of singularity in the pressure solution occur at the ends of intersection lines and high gradients in the pressure solution and flow fields occur close to the intersections. 
To properly resolve  these gradients the mesh needs to be finer in these regions. 
If fracture properties are homogeneous within a fracture (uniform fracture apertures) or only a pressure solution is required, or transport will be simulated using particle tracking then the mesh can be coarsened away from the intersections without significant loss of accuracy. 
However, if non-uniform apertures are considered, as in~\citep{dreuzy2012influence,makedonska2016evaluating}, then the mesh needs to be sufficiently fine  that length scales in the aperture field, such as  correlation lengths, are resolved.
Furthermore,  if transport is simulated using an Eulerian approach via a numerical discretization of the advective-dispersion equation where numerical diffusion/dispersion is controlled by the mesh resolution then a uniform mesh is more appropriate because numerical errors will be uniform across the domain. 

\begin{figure}[htb!] \centerline{
 \begin{tabular}{ccc}
(a) &(b) \\
 \includegraphics[width=0.45\textwidth]{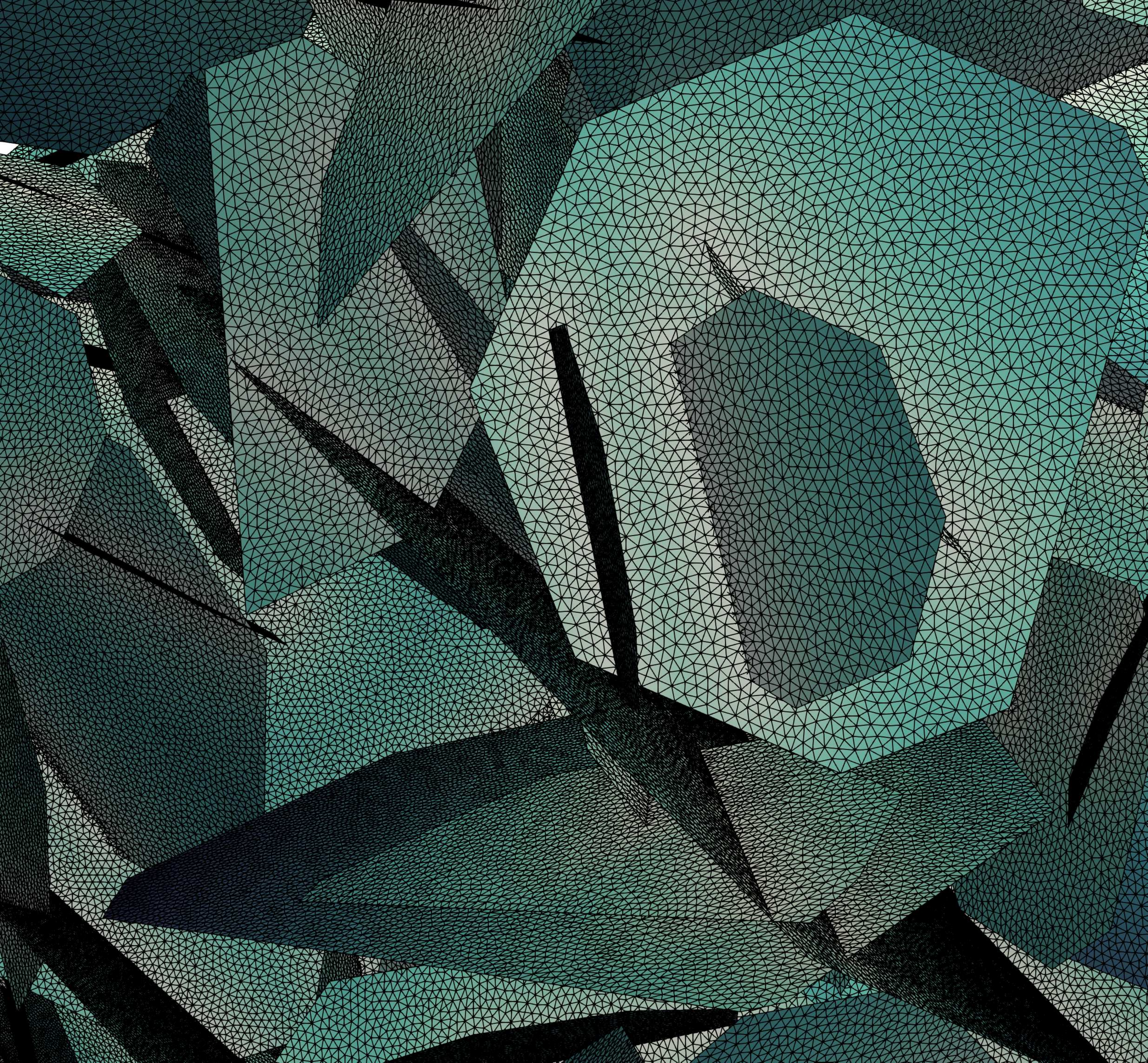}  &
 \includegraphics[width=0.45\textwidth]{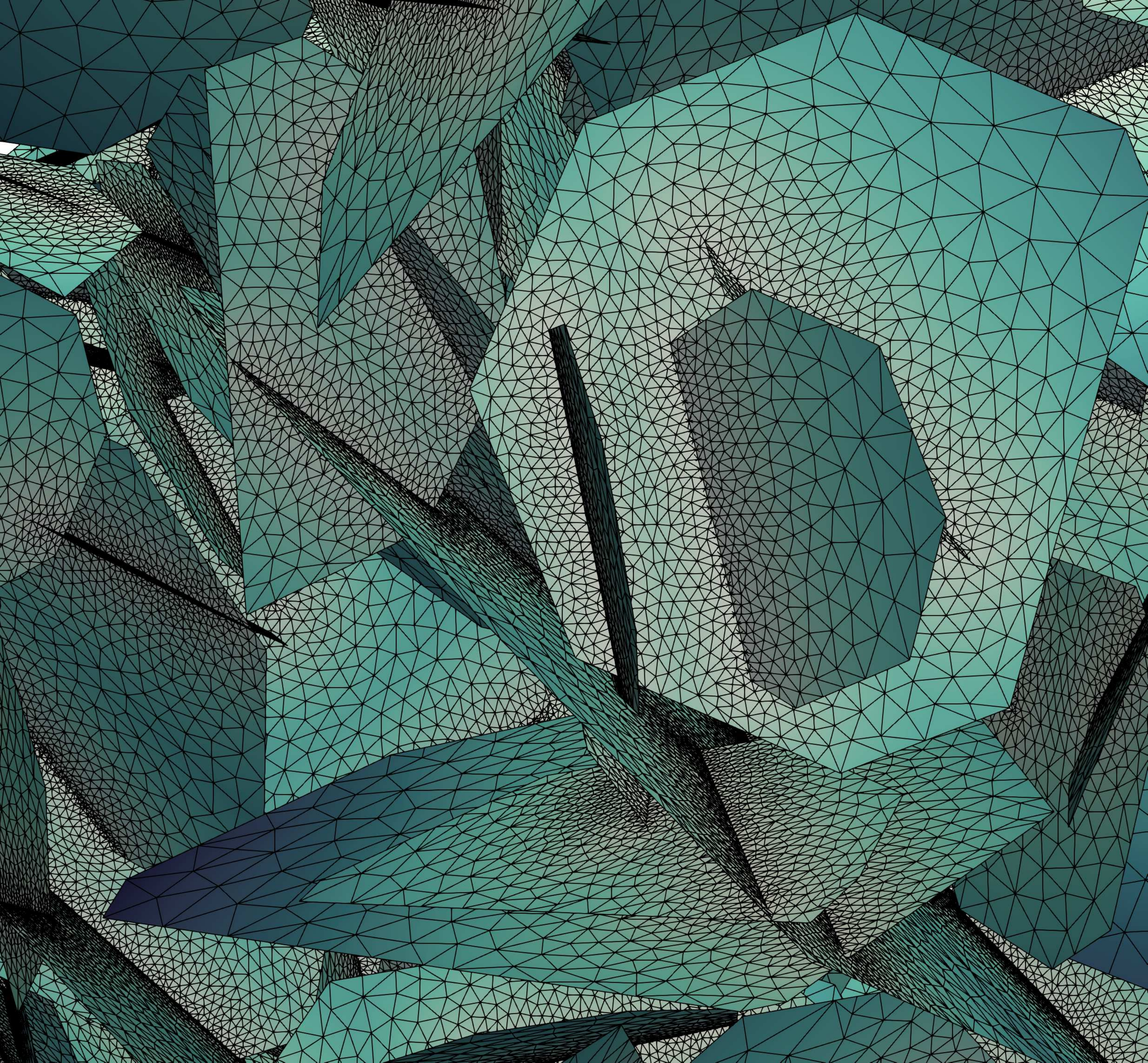}  \\
 \end{tabular} }
\caption{\label{fig:dfn_mesh} (a) Close view of uniform mesh resolution for the DFN shown in Fig.~\ref{fig:dfn}. (b) Close view of variable mesh resolution. In (b) the mesh is  coarsened away from fracture intersections to reduce the overall size of the mesh. The mesh shown in (a) is composed of 870,685 nodes and 1,712,924 triangles while the mesh in (b) is made up of 360,912 nodes and 725,787 triangles}
\end{figure}

Figure~\ref{fig:dfn_mesh}~(a) provides a close up view of uniform mesh resolution on the network shown in Fig.~\ref{fig:dfn} and  Fig.~\ref{fig:dfn_mesh}~(b) shows a close view of variable mesh resolution in the same region.  
In Fig.~\ref{fig:dfn_mesh}~(b), the mesh is coarsened away from fracture intersections to reduce the overall size of the mesh using the method described above. 
The mesh shown in Fig.~\ref{fig:dfn_mesh}~(a) is composed of 870,685 nodes and 1,712,924 triangles while the mesh in (b) is made up of 360,912 nodes and 725,787 triangles. 
In our development of a mesh partitioning method, we consider both uniform and variable resolution meshes.


\section{Graph Partitioning} 

Large DFNs with $\mathcal{O}(10^4)$ fractures, can have meshes with $\mathcal{O}(10^6)$ nodes even using variable resolution methods~\citep{makedonska2018discrete}.
Simulating flow and transport on these networks requires solving a sparse linear system for pressure $Ax=b$ by either direct solvers or  iterative methods.
In computations using multiple processors, one wants to minimize the communication between processors and evenly balance the work performed on each processor.
This problem of minimizing communication and load balancing is identical to the problem of partitioning the graph corresponding to the sparsity pattern of matrix $A$~\citep{kumar1994introduction}, which in our problem, is equivalent to partitioning the mesh of the DFN.
Thus, for a computer with $k$ processors, we seek a partition of the graph based on the DFN mesh into $k$ parts of equal size where the edges between those parts is minimized. 

\subsection{\emph{k-way} Graph Partitioning}

Formally, given a graph $G=(V,E)$ composed of vertices $u \in V$ and edges $e_{i,j} = e(u_i,u_j) \in E$, with non-negative vertex weights $w_i: V \to \mathbb{R}^+$ and edge weights, $w_{i,j}:E \to \mathbb{R}^+$\, let  $P = (P_1, \dots, P_k)$, be a partition of the vertex set $V$ into $k$ parts such that,
\begin{equation}
        \cup_i P_i  = V,
\end{equation}
and
\begin{equation}
        P_i \cap P_j = \emptyset \quad \text{for } i \neq j~.
\end{equation}
For a given partition we can measure the volume of each piece of the partition
\begin{equation}
|P_j| := \sum_{u_i \in P_j} w_i.
\end{equation}
The volume of each piece of the partition is used to provide a measure of imbalance.
For an imbalance parameter $\epsilon >0$, we can determine if $P$ satisfies the balance constraint
\begin{equation}\label{eq:balance}
\max_i |P_i | \leq (1 + \epsilon) \left \lceil \frac{|W |}{k} \right \rceil~,
\end{equation}
where $|W| = \sum_{u_i \in V} w_i$. 
Moreover, we can also measure the cut of a partition
\begin{equation}\label{eq:cut}
C(P) = \sum w_{i,j} \text{ s.t. } e_{ij} \in E, u_i \in P_k, u_j \in P_l   \text{ and }   k \neq l.
\end{equation}
The \emph{k-way graph-partitioning problem} (GP) is to find a $k$-partition, $P$, that satisfies the balance constraint (\ref{eq:balance}) and minimizes the cut (\ref{eq:cut}). 
In general, these two requirements conflict with one another.
Indeed, this graph partition problem is an NP-hard problem \citep{garey1974some,hyafil1973graph}.

\subsection{ Multilevel Graph Partitioning}

Multilevel graph partitioning is one of the most successful heuristics for partitioning large graphs~\citep{bui1993heuristic,cheng1991improved,garbers1990finding,hagen1992new,hendrickson1995multi,karypis1998fast,mansour1993graph}. 
The idea behind multilevel graph partitioning originates from the multiscale optimization and multigrid strategies~\citep{brandt2003multigrid}.
A graph is gradually coarsened to one where a  \emph{k-way} partition can be computed efficiently and effectively and then this partition is projected back onto the original graph.
To be more specific, let us consider a weighted graph $G_0 = (V_0, E_0)$ that has weights on both vertices and edges.
Algorithm~\ref{alg:multilevel} summarizes the multilevel framework for graph partitioning.

\begin{algorithm}
\caption{Multilevel Graph Partitioning}
\label{alg:multilevel}
\begin{algorithmic}
\Require: $G_0 = (V_0,E_0)$ with vertex weights $w_i$ and edge weights $w_{i,j}.$
\Ensure: $P(G)$
\begin{enumerate}
\item Coarsening phase : The graph $G_0$ is transformed into a sequence of smaller graphs $G_1$, $G_2$, \ldots, $G_m$ such that $|V_0| > |V_1|   > |V_2|  > \ldots |V_m|   $.
\item Initial (coarsest graph) partitioning phase: a high-quality algorithm is employed to obtain a \emph{k}-way partition $P_m$ of the graph $G_m = (V_m,E_m)$.
\item Uncoarsening phase: The partition $P_m$ of $G_m$ is projected back to $G_0$ via the intermediate partitions $P_{m-1},P_{m-2} \ldots, P_{1},P_{0}$ which are refined at each level $l\in [0,..,m-1]$.
\end{enumerate}
\end{algorithmic}
\end{algorithm}

The approach consists of three main phases: (I) coarsening, (II) initial partitioning and (III) uncoarsening.
In the \textit{coarsening phase} the original graph ($G_0$) is gradually approximated by creating a hierarchy of coarsened graphs, $G_1$, $G_2$, \ldots, $G_m$, where there is a decreasing number of vertices in each graph $|V_0| > |V_1|   > |V_2|  > \ldots |V_m|$ and subscript denotes partitioning level.
This can be achieved by collapsing edges and creating coarse level vertices, which are the nodes in the next level of the hierarchy that represent sets of vertices in next-coarser levels. 
The coarsening phase is stopped when the graph is small enough to be partitioned using an expensive but accurate algorithm. 
This phase is referred to as the \textit{initial partitioning phase}.
After the initial partitioning is performed, the \emph{uncoarsening phase} begins, which is made up of two parts.
In the first part of this stage, the partition at the coarser level $P_i$ is projected onto the graph one level finer in the hierarchy $G_{i-1}$, $P_i \rightarrow P_{i-1}$.
Next, this projected partition is refined using a variant of standard improvement algorithms to create a better partition at this level in the hierarchy, cf.~\citep{bulucc2016recent}.
This is done until $P_0$ is obtained. There are other (sometimes more sophisticated) multilevel frameworks for partitioning \citep{meyerhenke2009new,safro2015advanced} and other cut-based problems on graphs such as the minimum linear arrangement \citep{safro2006graph}, wavefront \citep{hu2001multilevel}, bandwidth \citep{safro2009multilevel}, and vertex separators \citep{hager2018multilevel}. 


\section{DFN-based Graph Partitioning}\label{sec:graph}

In this section, we describe one of the most common graph-representations of a DFN and develop methods to use that graph-representation in the partitioning of the mesh. 
We adopt a graph representation of a meshed DFN defined as a tuple $(\mathcal{F},I)$ where vertices in the graph correspond to elements in the set of fractures $\mathcal{F}$ and edges correspond to elements in the set of intersections $I$.
Formally, let  $\mathcal{F} = \{f_i \}$ for $i=1,\ldots, N$ denote a fracture network composed of $N$ fractures ($f_i$). 
Every $f_i \in \mathcal{F}$ is assigned a shape, location, and orientation within the domain by sampling distributions whose parameters are determined by a site characterization.
Every $f_i \in \mathbb{R}^2$ but the network $\mathcal{F} \in \mathbb{R}^3$. 
Let $I = \{(f_i, f_j)\}$  be a set of pairs associated with intersections between fractures; if $f_i  \cap f_j \ne \emptyset$ then $(f_i, f_j) \in I$.
The number of intersections depends on the particular shape, orientation, and geometry of the set of fractures in the network.
We denote the line of intersection between $f_i$ and $f_j$ as $\ell(f_i, f_j)$.
Using these sets, the topology of a DFN can be defined as the tuple $(\mathcal{F},I)$.
Hyman et al.~\cite{hyman2018identifying} recently showed that this particular graph-representation of a DFN is a projection of a more general bi-partite graph.

A simple undirected graph $F=(V_F,E_F)$ is constructed in the following way using an operator applied to the tuple $(\mathcal{F},I)$, $\phi:(\mathcal{F},I) \rightarrow (V_F,E_F)$.
For every $f_i \in \mathcal{F} $, there is a unique vertex $u_i \in V_F$, 
\begin{equation}\label{eq:phi1}
\phi_\mathcal{F} : f_i \rightarrow u_i \,.
\end{equation}
The vertex weight $w_i$  for vertex $u_i \in V_F$ is the number of mesh nodes on the fracture $f_i$, denoted $M_f$,
\begin{equation}\label{eq:vertex_weight}
w_i = M_f (f_i)~,
\end{equation}
Edges are defined in the following way. 
If two fractures, $f_i$ and $f_j$ intersect, $(f_i,f_j) \in I$, then there is an edge in $E$ connecting the corresponding vertices, 
\begin{equation}\label{eq:phi2}
\phi_I :  (f_i,f_j) \in I \rightarrow e_{ij} = e(u_i,u_j)\,,
\end{equation}
where $(u,v)\in E_F$ denotes an edge between vertices $u$ and $v$.
The weight $w_{i,j}$ of the edge $e(u_i,u_j) \in E_F$ is the number of mesh nodes on the intersection $(f_i,f_j) \in I$, denoted $M_I$,
\begin{equation}\label{eq:edge_weight}
w_{i,j} = M_I [(f_i,f_j)]~,
\end{equation}
This particular mapping has been used by a variety of researchers~\citep{aldrich2017analysis,andresen2013topology,hope2015topological,huseby1997geometry,hyman2017predictions,hyman2018dispersion,srinivasan2018quantifying,valera2018machine,viswanathan2018advancing}.
Note that this mapping of a DFN readily accounts for intersections of intersections~\citep{hyman2018identifying}, however such features require attention when meshing~\citep{hyman2014conforming}.

\begin{figure}[htp]
	\begin{center}
		\begin{tabular}{ll}
	 		(a) &(b) \\ 
			\includegraphics[width=0.45\linewidth]{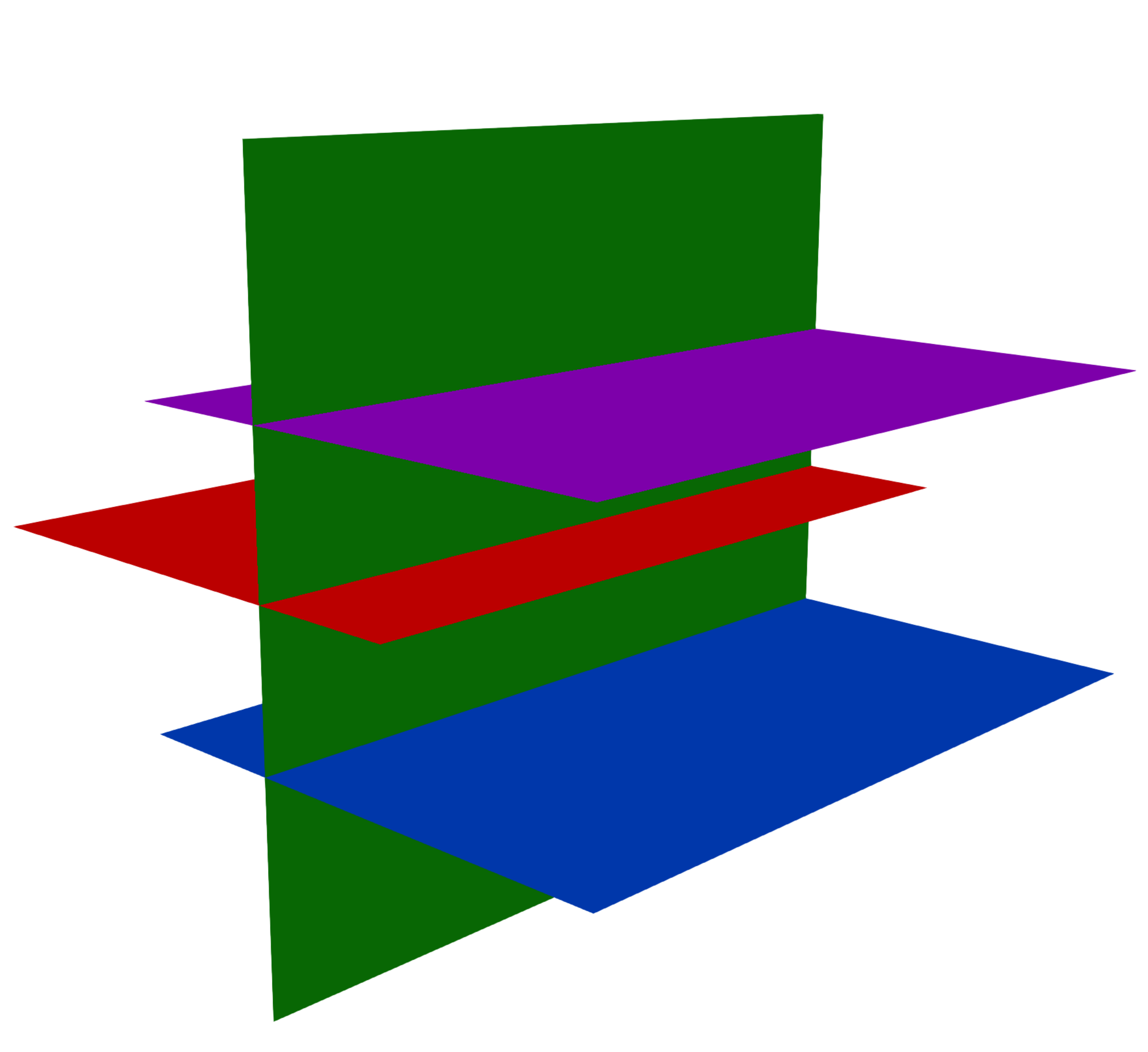} &  \includegraphics[width=0.45\linewidth]{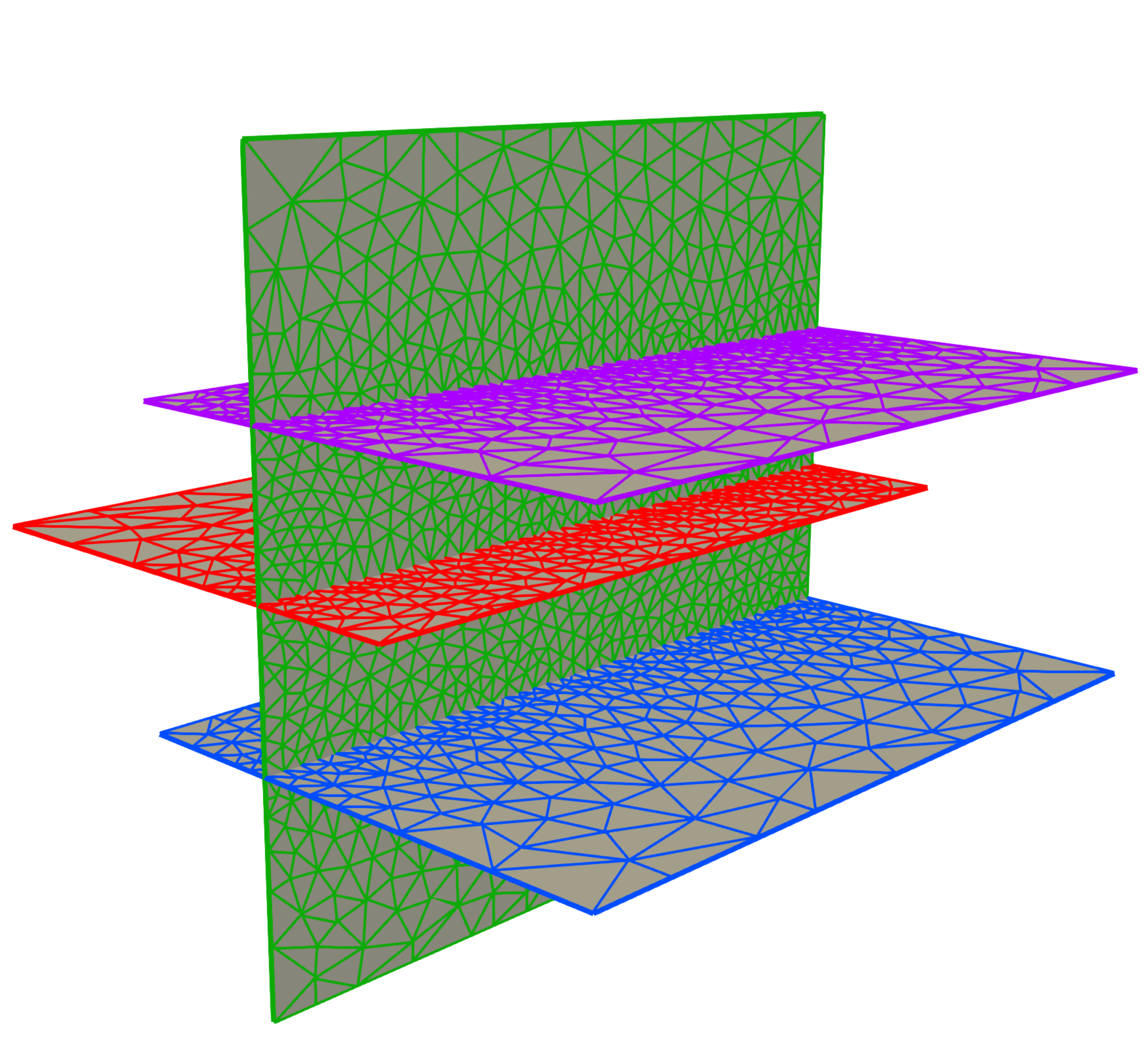} \\
			(c) & (d)  \\
			\includegraphics[width=0.45\linewidth]{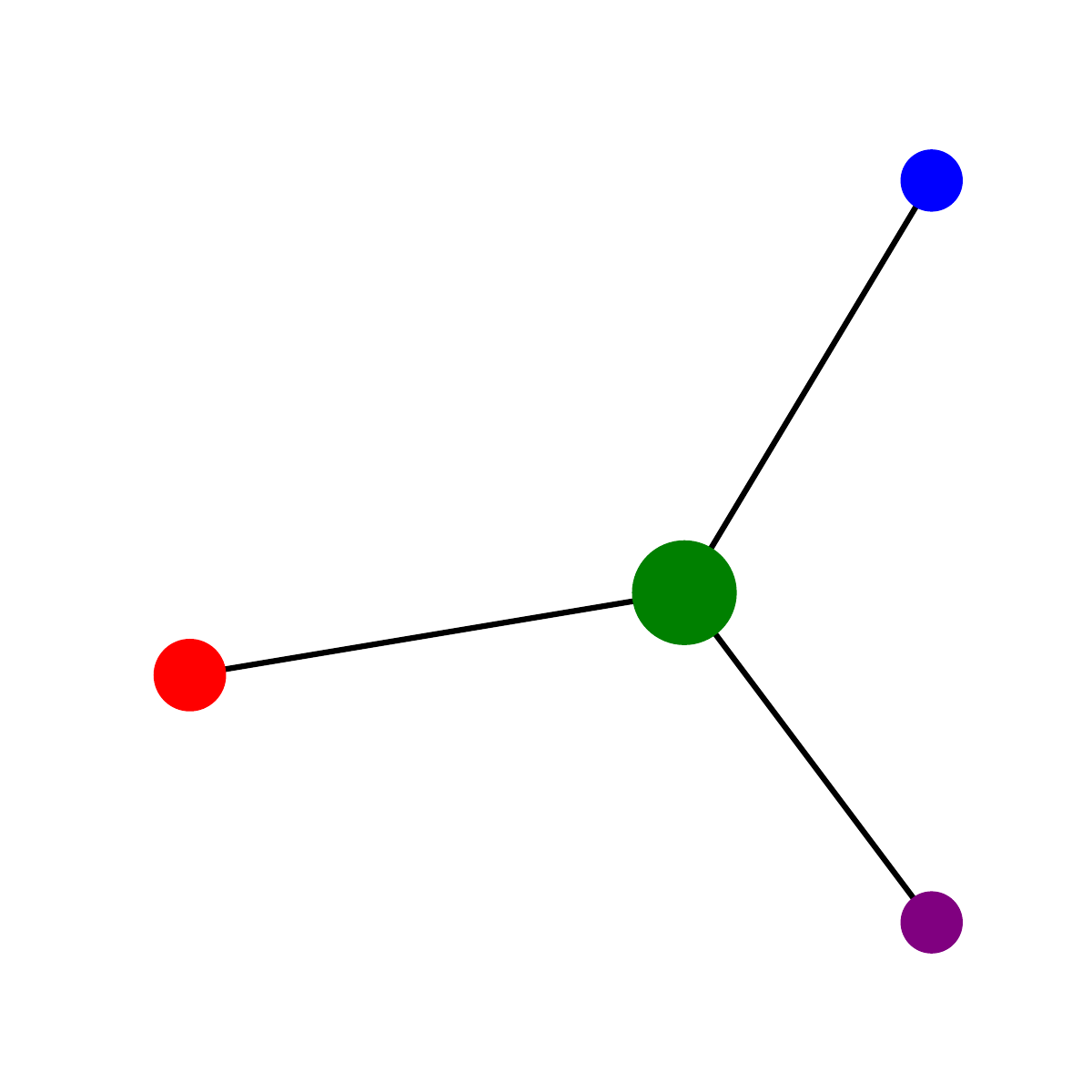} & \includegraphics[width=0.45\linewidth]{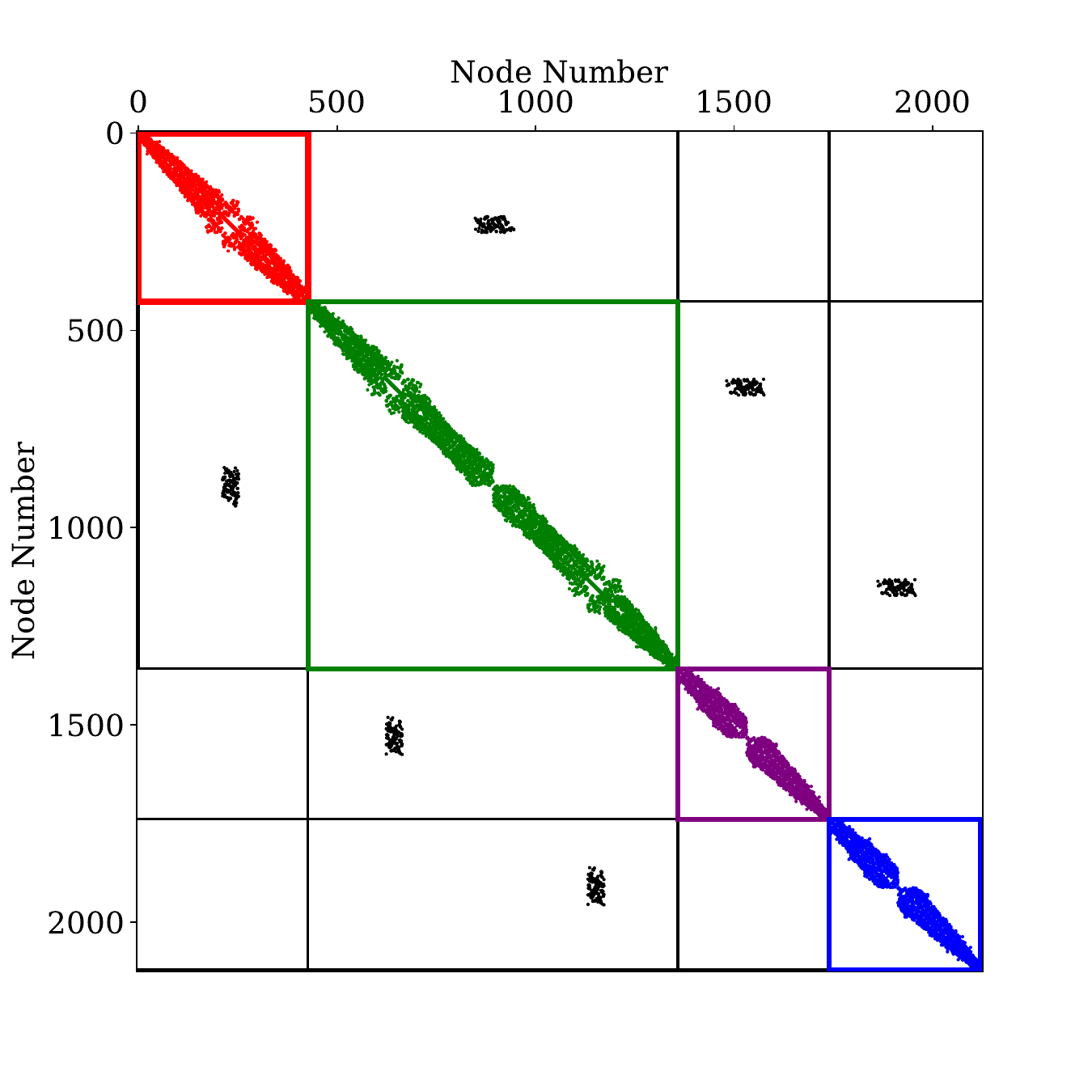} \\
		\end{tabular}
	\end{center}
\caption{\label{fig:four_fractures} (a) A DFN composed of four fractures. (b) The computational mesh on the DFN. Mesh colors coincide with the fracture on which the mesh node resides. (c) A graph representation of the DFN where vertex colors correspond with the fracture colors in (a). (d) The adjacency matrix for the mesh shown in (c). Colors in the matrix correspond to the fractures on which the nodes reside; black entries correspond to mesh nodes along fracture intersections}
\end{figure}

Figure~\ref{fig:four_fractures} shows a DFN composed of four fractures to demonstrate the connection between the graph-representation and the mesh. 
Figure~\ref{fig:four_fractures}(a) shows the DFN where each fracture has a unique color. 
Figure~\ref{fig:four_fractures}(b) shows the DFN with the mesh overlaid on the DFN, where the mesh colors correspond to the fracture on which they reside. 
Figure~\ref{fig:four_fractures}(c)  shows the adopted graph-representation of the DFN where vertex colors coincide with the fracture colors and vertex size corresponds to the vertex weight. 
Figure~\ref{fig:four_fractures}(d) is a plot of the adjacency matrix of graph equivalent of the mesh where colors in the matrix correspond to the fractures on which the nodes reside. 
We perform a multi-index  sort of the mesh nodes -- first by fracture number, then x coordinate, y coordinate, and finally z coordinate. 
This sort reduces the bandwidth of the main diagonal of the adjacency matrix. 
The block structure of the mesh is a direct result of the fracture network topology, which is captured in the graph plot in Fig.~\ref{fig:four_fractures}(c). 
The mesh nodes on each fracture make up the main diagonal of the adjacency matrix in the plot shown in Fig.~\ref{fig:four_fractures}(d).
The off diagonal nodes (black) correspond to mesh nodes along the fracture intersections. 
Each of these blocks corresponds to a single vertex in graph shown in Fig.~\ref{fig:four_fractures}(c) and the number of non-zero entries in each block corresponds to the weight of the vertex.
Mesh connections are mostly on a single fracture and there are fewer connections across fracture intersections, as indicated by the few off-diagonal terms in the adjacency matrix.


\subsection{Multilevel DFN-based Graph Partitioning}
We now propose an application of the multilevel graph partitioning algorithm that takes advantage of the topology of a DFN. 
The basic idea behind the method is to perform the partitioning on a graph based on the topology of the DFN and then projecting the resulting partition onto the DFN mesh.
In a variant of the method, this projection is used as the initial condition for the mesh partitioning as a final step in the algorithm.

Hyman et al.~\citep{hyman2018identifying} showed that the graph representation  $F$ defined by equations (\ref{eq:phi1}) and (\ref{eq:phi2})   is isomorphic to a DFN $\mathcal{F}$. 
An implication of that is that for every partition of the graph based on the DFN $P(F)$, there is a corresponding unique partition of the DFN $P(\mathcal{F})$.
This follows directly from the properties of the mapping  $\phi$ being a bijection. Applying $\phi^{-1}$ to $P(F)$ defines a unique $P(\mathcal{F})$. 
Therefore, we can partition a DFN using this graph representation.
However, we seek to partition the mesh of the DFN, not just the DFN.
Let $G=(V_G,E_G)$  be the graph defined by the conforming Delaunay triangulation of the DFN.
Note that with the exception of nodes along the lines of intersection in the DFN, every vertex $v \in V_G$ corresponds to a node in the mesh that resides on a single fracture $f_i \in F$. 
Let $f(v) = f_i$ be a function that returns the fracture on which the node corresponding to the vertex $v$ resides.  
For nodes on intersections between multiple fractures $f_i$ and $f_j$, let $f(v) = \min(f_i,f_j)$ where the minimum is taken over the fracture index.

We define a mapping $\Pi: G \rightarrow F$ to the graph $F$
\begin{equation}\label{eq:dfn_coarse1}
\Pi: v \mapsto f(v), \text{ for } v \in V_G~.
\end{equation}
Define vertex weights on $w_v \in V_F$ as the number of nodes in the mesh that reside on each fracture, and the edge weights in $E_F$ by as the number of nodes along the lines of intersections between fractures. 
Note that $F$ is the graph defined according to equations (\ref{eq:phi1}) and (\ref{eq:phi2}), the graph based on the topology of the DFN where each vertex corresponds to a fracture and edges indicate that fractures intersect.   
We retain information about the number of vertices that each coarse node in $V_F$ represents by using \eqref{eq:vertex_weight} and \eqref{eq:edge_weight}.
The graph $F$ is  a coarse version of the mesh-based graph but $|V_F| \lll |V_G|$ by several orders of magnitude.
We can apply the standard multilevel graph-partitioning method to $F$ and obtain $P$ for a {\emph k-way} partition.
Conceptually, the proposed method defines the first level in the coarsening phase  $\Pi: G_0 \rightarrow G_1 \equiv F  $ and then a partition $P(F) $ is obtained using Algorithm~\ref{alg:multilevel}. 
Once the partition $P(F)$ is obtained, we project the partition onto $G$ using $\Pi^{-1}$.
Note that $\Pi$ is not injective, but we adopt this notation for consistency.
In other words, if a fracture $f_i \in P_j$, then all nodes in the mesh on $f_i$, $f(v) = f_i$, are placed into $P_j$ of $G$.  
Note that the projection of a partition $P_F$ of the graph $F = (V_F, E_F)$ defined by equations (\ref{eq:phi1}) and (\ref{eq:phi2}) onto graph based on the mesh of the DFN $G=(V_G,E_G)$ 
\begin{equation}
\Pi^{-1}:P(F) \rightarrow P(G)~,
\end{equation}
 is unique. 
By definition, every $v \in V_F$ is in a unique part of the partition $P$ and equation~(\ref{eq:dfn_coarse1}) is surjective. 
Therefore all $v \in V_G$ in the pre-image of $v \in V_F$ are in a unique part of the partition $P$. 

Algorithm \ref{alg:multilevel_DFN} summarizes the method
\begin{algorithm}
\caption{Multilevel Graph Partitioning For DFN}
\label{alg:multilevel_DFN}
\begin{algorithmic}
\Require $F = (V_F,E_F)$ \Comment Graph based on DFN
\Ensure $P(G)$ \Comment Partition of the mesh of the DFN
\State $F_0 = F$ \Comment Initialize Multilevel method with finest level being the DFN based graph 
\State Perform Algorithm~\ref{alg:multilevel} on F\\
\begin{enumerate}
\item Coarsening phase : The graph $F_0$ is transformed into a sequence of smaller graphs $F_1$, $F_2$, \ldots, $F_m$ such that $|V_0| > |V_1|   > |V_2|  > \ldots |V_m|   $ 
\item Initial partitioning phase: A local refinement algorithm is employed to obtain a \emph{k}-way partition $P_m$ of the graph $F_m = (V_m,E_m)$
\item Uncoarsening phase: The partition $P_m$ of $F_m$ is projected back to $F_0$ via the intermediate partitions $P_{m-1},P_{m-2} \ldots, P_{1},P_{0}$ with subsequent refinements
\end{enumerate}
\\
\State $\Pi^{-1}: P_0(F_0) \rightarrow P_0(G)$ \Comment Project the partition of $F_0$ onto  the mesh of the DFN $G$
\end{algorithmic}
\end{algorithm}

The proposed procedure drastically simplifies the coarsening phase because it reduces the number of steps that need to be taken to reach a graph $F_m$ where a \emph{k-way} partition can be obtained, because the difference in size between $G$ and $F$ is large. 
Moreover, it reduces the complexity of the uncoarsening phase, because $P$ only needs to be obtained on $F$, not $G$. 
In practice, the mesh G is never constructed explicitly, only $F$ needs to be passed to the multilevel GP and the solution passed to the mesh.
Conceptually, Algorithm~\ref{alg:multilevel_DFN} skips many of the levels in Algorithm~\ref{alg:multilevel} and is essentially a two-level approach.

\begin{figure}[htp]
	\begin{center}
	 	\includegraphics[width=0.45\textwidth]{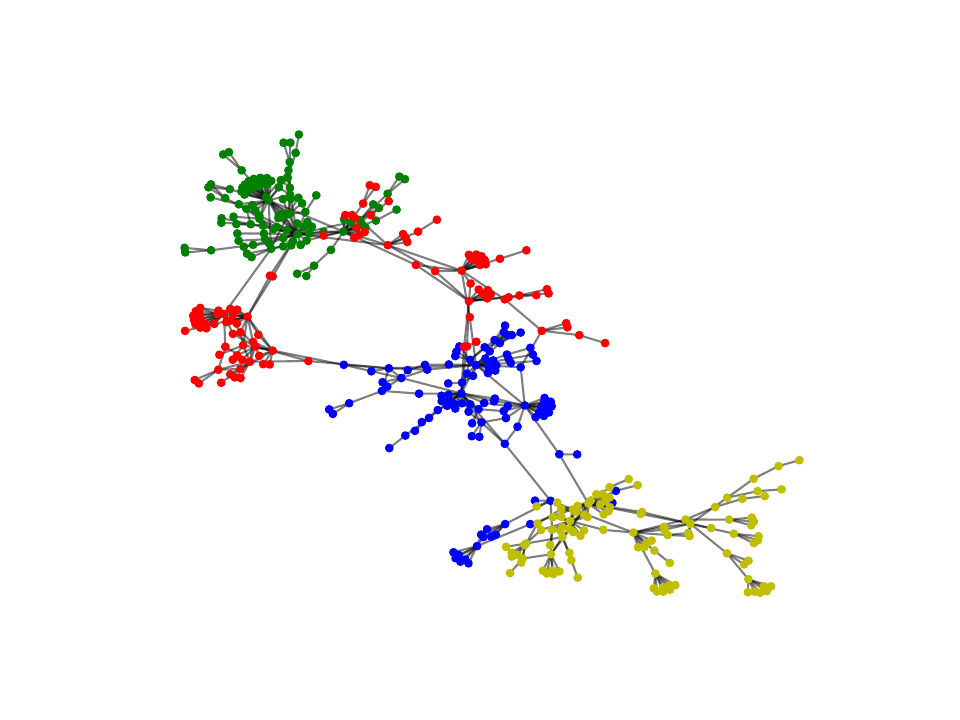} \\
	 	\includegraphics[width=0.45\textwidth]{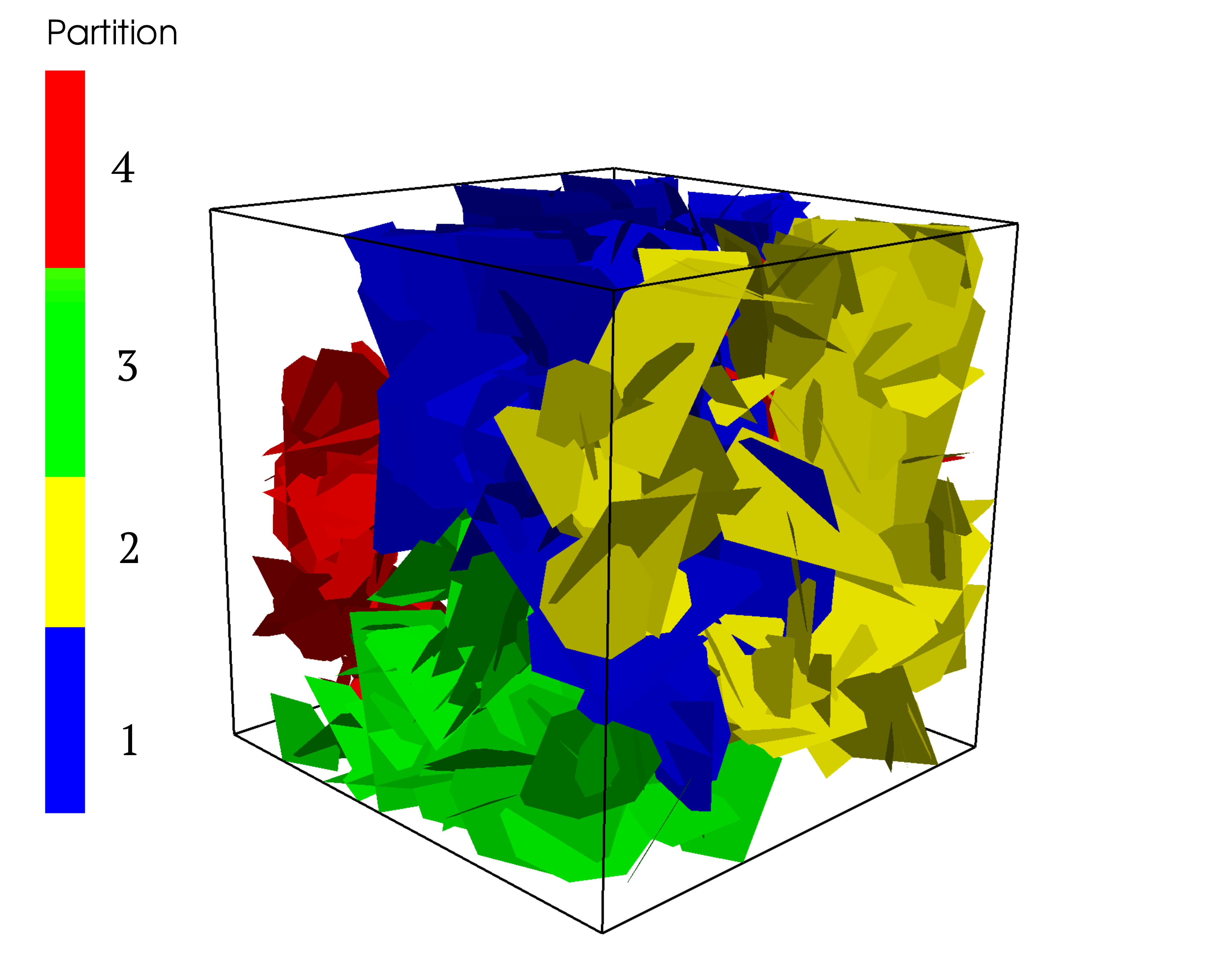}  \\ 
	\end{center}
\caption{\label{fig:partition} (Top) Graph-representation of the topology of the DFN. (Bottom) DFN colored based on a four-part DFN-based partition. The partition of the DFN-mesh (not-shown) is obtained by projecting the partition of the DFN-based graph onto the mesh}
\end{figure}

As an example, the DFN shown in Fig.~\ref{fig:dfn} is made up of 424 fractures, so the graph-representation has 424 nodes, while the mesh has 870,685 nodes for the uniform mesh and 360,912 nodes for the variable resolution. 
Figure~\ref{fig:partition} (top) shows the graph based on that fracture network $F$ colored according to a four-way partition. 
The DFN is shown on the bottom of the image, where  colors correspond to the partitions in  $F$ (the DFN is colored by $P(G)$).
The projected partition requires that cuts occur along intersections, meaning that intersections cannot divided by a cut. 
The degrading colors in the figure within a fracture plane are a result of the position of the light source in rendering the image.
Note that the projection $\Pi^{-1}$ to obtain the partition $P(G)$ is agnostic to the meshing strategy and resolution. 
However, the meshing strategy does affect the quality of the cut in the projected partition $P(G)$ and we present a slight modification to the method in the next subsection.


\subsection{Multilevel DFN-based Graph Partitioning with Refinement}
One undesirable feature of Algorithm \ref{alg:multilevel_DFN} is that the projected cuts obtained from the DFN graph onto the mesh are necessarily along lines of intersection in the mesh. 
If a variable mesh resolution is used, these regions have the highest density of mesh elements and the projected cuts can be quite large. 
To address this issue, we add a final step of the algorithm where the projected cut obtained using the DFN graph, is used as the initial condition for the mesh partitioning. 
Thus, it is a hybrid method capitalizing on the strengths of both partitioning the mesh directly and partitioning the coarse topological representation.
The modified algorithm is detailed in Algorithm~\ref{alg:multilevel_DFN_refine}.

\begin{algorithm}
\caption{Multilevel Graph Partitioning For DFN with refinement }
\label{alg:multilevel_DFN_refine}
\begin{algorithmic}
\Require $F = (V_F,E_F)$ \Comment Graph based on DFN
\Ensure $P(G)$ \Comment Partition of the mesh of the DFN
\begin{enumerate}
	\item  $F_0 = F$ \Comment Initialize Multilevel method with finest level being the DFN based graph 
	\item  Perform Algorithm~\ref{alg:multilevel} on $F$
	\item  $\Pi^{-1}: P_0(F_0) \rightarrow P_0(G)$ \Comment Project the partition of $F_0$ onto  the mesh of the DFN $G$
	\item  Perform Algorithm \ref{alg:multilevel} with $P_0(G)$ as the initial condition.
\end{enumerate}
\end{algorithmic}
\end{algorithm}

Once a partitioning of the DFN based graph $F$ is found, an initial partitioning of the mesh of the DFN $G=(V_G, E_G)$ is determined by assigning each node $v \in V_G$ to the part assigned to the node $\Pi(v) \in V_F$. This initial partitioning is then improved in either of the following ways: (I) with a refinement (improvement based) algorithm that iteratively improves on the initial solution until a convergence criteria is achieved, or (II) the initial partioning is passed on to a multilevel solver that takes the initial solution into account within the coarsening phase by only merging nodes within the same part. Both of these methods have the potential to improve on the solution, with the later method often achieving higher quality solutions with potentially a longer running time. 

Given that the size of the graph $F$ is orders of magnitude smaller than $G$, we obtain a good partition quickly and then refine it. 
In practice, this results in a partition almost as good as directly partitioning $G$ but at a fraction of the computational cost. 
In the next section, we provide a suite of examples to compare these algorithms and characterize their properties.
Given that the size of the graph $F$ is orders of magnitude smaller than $G$, we obtain a good partition quickly and then refine it. 
In practice, this results in a partition almost as good as directly partitioning $G$ but at a fraction of the computational cost. 
In the next section, we provide a suite of examples to compare these algorithms and characterize their properties.

\section{Numerical Examples}\label{sec:results}

We compare the proposed approach, where the partition of the mesh is based on the partition of the graph representation of the DFN, with the standard approach, where the mesh is partitioned directly.
We consider two sets of simulations to demonstrate the utility of the method. 
In the first, we consider a set of 30 independent identically distributed DFN realizations with both variable and uniform mesh resolution.
In DFN modeling, it is commonly required that an ensemble of network realizations are considered, rather than running a very large single simulation.
The focus of this case study is to understand the relative gains of considering an ensemble.  
The meshes of these networks are partitioned into 2, 4, 8, and 16 partitions. 
In the second case study, we consider a single large DFN containing 2,432 fractures.
This network is selected to be representative of the larger sized fully resolved DFN models currently being used.
For this network, we consider 128, 256, and 512 partitions.
The networks used in the two studies are substantially different and comparison is meant to be within the case studies, but not between them.
Generation and meshing of the fracture networks is performed using the {\sc dfnWorks} computational suite~\citep{hyman2015dfnworks}.
A conforming Delaunay triangulation on each network is performed using the feature-rejection algorithm for meshing ({\sc fram})~\citep{hyman2014conforming}.  
The parallelized subsurface flow and reactive transport code {\sc pflotran}~\citep{lichtner2015pflotran}, which uses the {\sc PETSc}~\citep{balay2017petsc} toolkit, is applied to solve the Laplace's equation for the steady-state volumetric flow rate and distribution of pressure in the network.
Note that this is a standard procedure in many applications in subsurface hydrology---solving a linear system for pressure in the network and then simulating transport using a particle tracking based approach.

For each network, we consider three partitions: (I) The partition obtained on the mesh itself (Algorithm~\ref{alg:multilevel}); we refer to these partitions as $P(G)$, (II) the partition induced from the partition on the graph representation of the DFN $P(F)$ (Algorithm~\ref{alg:multilevel_DFN}), and (III) the refined partition (Algorithm~\ref{alg:multilevel_DFN_refine}) $P(F/G)$. 
In our experimental results, we use the graph partitioning package KaHIP \citep{sanders2013think} which among other methods implements the Global Path algorithm for matching, and flow-based methods for partition refinement.
The quality of the partitions is judged by the cut (number of edges that link between partitions) and the imbalance (the difference in sizes of the partitions). 
Our choice in using KaHIP as baseline graph partitioning solver stems from the fact that KaHIP has been experimentally shown to produce high-quality  partitions for graphs that exhibit a power-law degree distribution, which is a property of the DFN graphs in this work~\citep{hyman2019linking,hyman2019emergence}.
For the DFN partitioning, we use the kaffpaE solver with the strong social pre-configuration and use the fast pre-configuration for the mesh partitioning. We choose the kaffpaE solver together with the strong-social pre-configuration mainly due to the fact the DFN graphs have a power-law degree distribution, similar to a social network that these methods are designed for. 
We also compare the impact of the partitions on computational performance by solving porous media flow equations, which are Laplace's equation under steady-state conditions, and solve for the distribution of pressure within the network.
Here, we compare the number of FLOPS, the wall-clock run time, and number of iterations to obtain the solution using a bi-conjugate gradient scheme with a block Jacobi  preconditioner using the {\sc PETSc}~\citep{balay2017petsc} toolkit.

\subsection{Ensemble of Networks}

Each DFN is constructed  in a cubic domain with sides of length \SI{15}{\meter} and are composed of circular fractures with uniformly random orientations and uniformly random centers.
Fracture radii $r$ [m] are sampled from a truncated power law distribution (a commonly observed property in the natural world~\citep{bonnet2001scaling}) with exponent $\alpha= 2.6$ and upper and lower cutoffs ($r_u= \SI{5}{\meter}$; $r_0=\SI{1}{m}$), with probability density function of
 \begin{equation}\label{eq:powerlaw}
p_r(r) = \frac{\alpha}{r_0} \frac{(r/r_0)^{-1-\alpha}}{1 - (r_u/r_0)^{-\alpha}}.  
\end{equation}
The choice of exponent and cutoffs are selected such that no single fracture directly connects inflow and outflow boundaries. 
Variability in hydraulic properties is included into the network by correlating fracture apertures to their radii.
We use a positively correlated power-law relationship $b = 5.0\times 10^{-5} \sqrt{r}\,$.

On average, the networks contain around 470 fractures. 
In the graph representation, there are around 470 nodes and 645 edges.
When using a uniform mesh, there are, on average, one million nodes in the mesh (997,221) and nearly two million triangles (1,964,988). 
The graph based on the uniform mesh is made up of just under one million vertices and close to 3 million edges (2,962,302), on average.
Thus,  when partitioning the graph based on the uniform mesh, there are two thousand times more vertices than when partitioning the graph based solely on the DFN topology. 
In the case of the variable mesh, there are around half a million nodes (415,206) and three quarter million triangles (836,452), on average. 
Therefore the graph based on the variable mesh is made up of just under half a million vertices and over one million edges (1,251,751), on average. 
Thus, when partitioning the graph based on the variable mesh, there are about one thousand times more vertices than when partitioning the graph based solely on the DFN topology. 


We begin by reporting the quality of the partitions and computation time. 
Table~\ref{tbl:metrics} reports the cut, imbalance, and times for the uniform and variable mesh resolution. 
Reported values are the average of the thirty realizations. 
Columns correspond to each partition and row are sorted by the number of partitions $k$. 
For the uniform mesh case,  the lowest cuts are all obtained for $P(G)$ for all values of $k$.
The cut values obtained for $P(F)$ are about twice as large as those obtained using $P(G)$ but partitioning $P(G)$ take four orders of magnitude longer than  partitioning $P(F)$.
The observed difference in cut values for $P(F)$ between uniform mesh and variable mesh is due to the different vertex weights in the DFN-based graph, due to different meshes, which results in slightly different partitions. 
In all cases, the imbalance values are about the same.
The cuts on $P(F/G)$ are about 30\% larger than $P(G)$, but require three orders of magnitude less time than those obtained for $P(G)$. 
Similar observations are made in the variable mesh case, but there are a few subtle differences. 
The difference in the partition quality in terms of the cut between $P(G)$ and $P(F)$ is substantially larger than in the uniform mesh resolution set. 
In the case of $k=16$, the cut  for $P(F)$ is three times larger than for $P(G)$.
This increase in the cut values is a result of the fact that cuts in $P(F)$ can only occur along intersections in the fracture network mesh, where the mesh is most refined and the highest number of nodes exists.
In contrast, $P(G)$ is not constrained in this manner and can therefore partition the mesh in region of the fracture where the mesh is coarse and fewer edges exists.
However, in the case of $P(F/G)$, the partition is much improved when compared to $P(F)$ at little extra run time. 
Here, $P(F)$ is the initial condition for $P(G)$, but a fast partition, opposed to strong, is obtained. 
A result of this is smaller times required to get a higher quality partition. 
Note that the times required for $P(G)$ compared to $P(F)$ and $P(F/G)$ are four to five orders of magnitude different, regardless of the number of processors.
All imbalance values are approximately the same.

\begin{table}[htp]
\caption{Partition Metrics}
\begin{center}
\begin{tabular}{ll|c|c|c||c|c|c|c}
 & & \multicolumn{3}{|c|}{Uniform Mesh} &\multicolumn{2}{|c|}{Variable Mesh} \\ \hline\hline
Metric & $k$& $P(G)$ &  $P(F)$  & $P(F/G)$     &    $P(G)$ &  $P(F)$  & $P(F/G)$  \\ \hline \hline
Cut  & 2    & 404.73   &   671.67  & 532.57  &   237.20 &   676.37    &  349.93             \\
    & 4  & 941.87   &   1,657.30 &  1,269.27 &   569.03 &   1,632.00  & 757.23             \\
    & 8  & 1,871.33 &   3,412.50 &  2,487.40 & 1,118.30 &   3,403.23  & 1,401.90             \\
    & 16 & 3,587.00 &   7,503.43 &  4,995.73 & 2,119.90 &   7,637.17  & 2,827.90             \\ \hline
Imbalance & 2 & 0.03 &  0.02 &   0.02 &      0.02 &   0.02   & 0.02                         \\
 & 4 & 0.04 &   0.04 & 0.04 & 0.04 &   0.04       & 0.04  \\
 & 8 & 0.05 &   0.04 & 0.04 & 0.05 &   0.04       & 0.04 \\
 & 16 & 0.05 &   0.05 & 0.05 & 0.05 &   0.05       & 0.05 \\ \hline
Wall-Clock Time [sec] & 2 & 313.84 &   0.13 & 1.05 & 86.68 &   0.13 & 0.44  \\
& 4 & 375.59 &   0.18 & 1.54 & 93.41 &  0.18 & 0.64 \\
 & 8 & 515.05 &   0.25 & 1.64  & 114.40 &  0.25 & 0.57 \\
& 16 & 415.08 &   0.35 & 1.99 & 109.10 &  0.35 & 0.62 \\
\end{tabular}
\end{center}
\label{tbl:metrics}
\end{table}%


Table~\ref{tbl:computation} reports the number of GFlops, iterations required for the Krylov solver to converge, and wall-clock run time using the partitions on the 30 networks. 
For all values of $k$, the selected metrics for the partitions $P(G)$ and $P(F)$ are roughly the same. 
An interesting observation is that even though the cuts of $P(F)$ are three times larger than those of $P(G)$ in the case of the uniform mesh, the run times are only slightly larger. 
Due to the fewer degrees of freedom in the variable mesh than the uniform mesh, the number of FLOPS, iterations, and solve time are lower than those reported for the uniform mesh. 
In general, the FLOPS and number of iterations are comparable between $P(G)$, $P(F)$ and $P(F/G)$. 
However, the run times  for $P(F)$ are slightly slower than for $P(G)$. 
This slight slow down is  likely related to aforementioned issues with the constrained cut location of $P(F)$. 
Note that the run times of $P(F/G)$ are comparable to  $P(G)$.

\begin{table}[htp]
\caption{Computation Metrics}
\begin{center}
\begin{tabular}{ll|c|c|c||c|c|c|c} & & \multicolumn{3}{|c|}{Uniform Mesh} &\multicolumn{2}{|c|}{Variable Mesh} \\ \hline\hline
Metric & $k$& $P(G)$ &  $P(F)$  & $P(F/G)$     &    $P(G)$ &  $P(F)$  & $P(F/G)$  \\ \hline \hline
GFlops & 2 & 45.8 & 44.5 & 45.4 & 12.7 &   12.5 & 12.3 \\
& 4 & 22.6 &  22.4 & 22.9 &  6.29 &   6.37  & 6.41\\
& 8 & 11.3 &   11.6  & 11.4 & 3.24 &   3.21  & 3.26\\
  & 16 & 5.94 & 5.75 & 5.82 & 1.62 &   1.64  & 1.65 \\                           \hline
Iterations & 2 & 1,223.10 &  1,180.20 & 1,205.07 &  804.03 &   788.13 & 784.70 \\
& 4 & 1,188.97 &   1,175.33 & 1,203.93 &  784.97 &   796.27 & 803.40 \\
& 8 & 1,180.53 &   1,211.60 & 1,197.00 &  806.60 &   799.90 & 809.80  \\
& 16 & 1,236.47 &   1,195.17 & 1,215.67 & 800.87 &   812.83 & 818.57 \\ \hline
Time [sec] & 2 & 62.35 &   55.31 & 56.77 & 14.38 &   16.29 & 16.10  \\
& 4 & 34.94 &   34.98 & 35.20 & 8.45 &   10.10 & 9.96 \\
& 8 & 21.48 &   22.34  & 21.54 & 5.77 &   6.43  & 6.29 \\
& 16 & 15.64 &   14.73 & 15.13 & 4.08 &   4.45  & 4.31\\  \hline
\end{tabular}
\end{center}
\label{tbl:computation}
\end{table}%


Table~\ref{tbl:total_time} reports the total time taken for both the uniform and variable mesh partitions. 
In all cases, the slowest run times are reported for the $P(G)$, primarily due to the time required for the partition. 
Note this also drastically affects the scaling of the total run time with number of processors. 
The fastest times are reported for $P(F)$.
And the total run times for $P(F/G)$ are slightly larger than $P(F)$.

\begin{table}[htp]
\caption{Total Time [sec]}
\begin{center}
\begin{tabular}{l|c|c|c|c|c|c|c|c|c|}
& \multicolumn{3}{|c|}{Uniform Mesh} & \multicolumn{3}{|c|}{Variable Mesh} \\ \hline \hline
 $k$ & $P(G)$ &  $P(F)$& $P(F/G)$ & $P(G)$ &  $P(F)$& $P(F/G)$ \\ \hline \hline
 2 & 376.19 &   55.44  & 57.82 & 101.06 &  16.42 & 16.54  \\
 4 & 410.53 &   35.16  & 36.74 & 101.86  & 10.28 & 10.60    \\
 8 & 536.53 &   22.59  & 23.18 & 120.17 &  6.68 & 6.86 \\
 16 & 430.72 &  15.07 & 17.12 & 113.11 & 4.80 & 4.92 \\
\end{tabular}
\end{center}
\label{tbl:total_time}
\end{table}%

\subsection{Large Network Example}

To study the performance of the proposed method at larger scales, we consider a DFN containing 2,432 fractures, meshed with 39,143,090  nodes and 79,581,226 triangles. 
Thus, the difference in the size of the graphs the partitioner will have to consider is $\mathcal{O}(10^5)$. 
The network is composed of three families of fractures whose mean orientations are orthogonal to one another. 
The fracture radii are sampled from a truncated powerlaw \eqref{eq:powerlaw} with an exponent of 1.2, lower cut off of 1 meter and upper cut off of 100 meters. 
We selected these parameters to have a larger example with significantly different topological and geometric properties, higher connectivity and wider range of length scales, than the smaller networks. 
The mesh has variable resolution, and a value of $h = 0.1$ is used. 
The domain size is cube with sides of 500 meters.
The fractures are uniformly distributed throughout the domain.
Again, we solve Laplace's equation with variable permeability coefficients to obtain values of volumetric flow rates and pressure throughout the domain. 
We divide the DFN/mesh into 128, 256, and 512 partitions. 

Table~\ref{tbl:large_dfn_stats_partition} reports the cut, imbalance, and partition times for both the mesh $G$ and the graph based on the fracture network $F$. 
The cuts obtained using the mesh are consistently an order of magnitude less than those obtained using the DFN, which is expected given the high resolution of the mesh close to the intersections on which the projection from $P(F)$ must occur. 
As the number of cores increases, so does the imbalance of $P(F)$.
However, the time required to obtain $P(F)$ is consistently four orders of magnitude less than that required for $P(G)$ due to the stark contrast in the graph sizes.
In the case of 512 cores, there are members of the $P(F)$ partition that are not assigned nodes in the mesh because there are too few fractures to be distributed among the partition. 
Using the additional refinement method $P(F/G)$, there is a drastic improvement in the cut values and imbalance. 
The final values are much closer to those obtained using $P(G)$ but at a fraction of the computational cost. 

The different values for the cut and imbalance affect the number of FLOPS, iterations, and solver time, which are reported in Table~\ref{tbl:large_dfn_stats_computation}.
In the case of the $P(G)$, the high-quality partitions lead to better performance of the solver; near ideal scaling is observed in run times. 
However, the substantially larger cuts and higher imbalances in $P(F)$ lead to more communication between processors and an imbalanced workload.  This results in  longer solver convergence times  than what is required for $P(G)$.
For  256 processors, the large imbalance and cut values lead to a solver time that is double that of $P(G)$ and larger than that required for 128 processors.
The overall computational time, which is reported in Table~\ref{tbl:large_dfn_stats_time},  is still much less because the DFN based partitioning only takes a fraction of the time of $P(G)$. 
Because the 512 partition of $P(F)$ did not provide mesh elements for all 512 processors, we were unable to run the simulation and a comparison cannot be made. 
In the case of the $P(F/G)$, the improved cut and imbalance values offered by the additional refinement results in significantly lower run times for the solver compared to the $P(F)$. 
The solver times are close to those obtained using $P(G)$ and the overall run times are slightly larger than $P(F)$.

\begin{table}[htp]
\caption{Large Network : Partition Metrics}
\begin{center}
\begin{tabular}{ll|c|c|c|}
Metric & $k$& $P(G)$ &  $P(F)$  & $P(F/G)$    \\ \hline \hline
Cut  & 128    & 85,013   &   1,264,321  & 113,751  \\
    &  256 & 145,887  &   2,080,333 &  191,007\\
    & 512 & 245,839 &   2,693,132 & 567,955 \\ \hline
Imbalance & 128 & 0.05 &   0.35 & 0.05   \\
 & 256 & 0.05 &   1.09 & 0.05 \\
 & 512 & 0.05 &   3.58 & 0.03 \\ \hline
Wall-Clock Time [sec] & 128 & 34,340.00 & 3.11   & 92.67  \\
 & 256 & 17,045.30&   1.35  & 43.37 \\
& 512 & 17,665.10 &   1.72 & 240.50 \\ \hline 
\end{tabular}
\end{center}
\label{tbl:large_dfn_stats_partition}
\end{table}%

\begin{table}[htp]
\caption{Large Network : Computation Metrics}
\begin{center}
\begin{tabular}{ll|c|c|c|}
Metric & $k$& $P(G)$ &  $P(F)$  & $P(F/G)$    \\ \hline \hline
GFlops  & 128    & 2.49 $\cdot 10^5$  &   2.88 $\cdot 10^5$  & 3.25 $\cdot 10^5$   \\
    &  256 & 1.34 $\cdot 10^5$  &   3.87 $\cdot 10^5$  &  1.11 $\cdot 10^5$  \\
    & 512 & 5.31 $\cdot 10^4$ &   - & 7.83 $\cdot 10^4$  \\ \hline
Iterations & 128 & 10,369 &  10,281 & 13,574  \\
 & 256 & 11,193 & 16,218 & 9,226 \\
 & 512 & 8,821 &   - & 13,308\\ \hline
Wall-Clock Time [sec] & 128 & 581.33 &  605.08 & 751.58 \\
 & 256 & 300.84 & 633.21  & 250.36  \\
& 512 & 127.91 &   - & 192.41  \\ \hline 
\end{tabular}
\end{center}
\label{tbl:large_dfn_stats_computation}
\end{table}%

\begin{table}[htp]
\caption{Large Network : Total Wall-Clock Time [sec]}
\begin{center}
\begin{tabular}{c|c|c|c|c|}
$k$& $P(G)$ &  $P(F)$  & $P(F/G)$    \\ \hline \hline
 128  & 34,921.33 &  608.19  & 844.25  \\
 256 & 17,346.14 &  634.56 &  293.73 \\
 512 & 17,793.01 &   - & 432.91  \\ \hline
\end{tabular}
\end{center}
\label{tbl:large_dfn_stats_time}
\end{table}%

\section{Summary and Conclusions}\label{sec:remarks}

DFN modeling is a powerful tool to improve our understanding of how the multi-scale structure of fractured media influences flow and transport therein. 
However, the explicit representation of these fracture networks, which can contain length scales spanning several orders of magnitude, is computationally demanding. 
As the number of fractures in a DFN increases, so does the size of the mesh and the associated physical systems to model physical phenomena within the DFN. 
This increase in computational requirements is compounded by the inherent uncertainty in the subsurface that requires numerous realizations of a DFN be considered to bound system behavior. 
The combination of these facts requires that DFN models utilize efficient multiprocessing methodologies to accelerate system solving time.
Load balancing and minimizing communication between processors are critical factors in such methodologies.
Thus, a cornerstone in the use of multiple processors for DFN simulations is a high-quality partition of the mesh representation of the DFN. 

We have presented a topologically-based method for mesh partitioning in DFN simulations that utilizes the intrinsic multilevel nature of the fracture network. 
The method combines multilevel graph partitioning with a coarse-scale graph representation of the DFN to drastically improve the speed of obtaining a high-quality partition of the DFN mesh.
We partitioned the graph based on the DFN, rather than the mesh itself, and a partition of the mesh is obtained by projecting the DFN partition onto the mesh. 
This projection can then be used as the initial condition for the partitioning of the mesh, thereby greatly accelerating the partitioning time. 
The significant difference in size between the graph-based on the mesh and the graph-based on the DFN topology with the DFN based partition lead to a mesh-based partition that required a fraction of the time.
This improvement is even more pronounced when a final step of refinement is performed using the DFN based partition as an initial condition for mesh partitioning. 

We demonstrated the utility of the method by applying it to two test cases.
While these test cases are relatively simple compared to real-world applications in terms of boundary conditions and network complexity, they demonstrate the feasibility of the methods. 
Future work will address how the improved speed up from this algorithm is affected by more complex physics such as multiphase and transient flow, which are beyond the scope of this study.
We compared the proposed method to standard mesh-based partitioning in terms of graph-based metrics (cut, imbalance, time to obtain the partition), computational-based metrics (FLOPS, iterations, solver time), and total run time.
In both cases, the proposed methods outperformed standard mesh partitioning in terms of the total time required to perform a steady-state flow simulation using multiple processors. 
In the first test case, we considered a set of 30 three-dimensional discrete fracture networks composed of approximately five hundred fractures apiece meshed with around one million nodes. 
We considered uniform and variable mesh resolutions in this ensemble of networks. 
Foremost, the time required to obtain the partitions using the graph based on the DFN topology is negligible compared to the time to get the partition of the mesh due to the drastic difference in the size of the corresponding graphs. 
Additionally, the time required for further refinement was insignificant, but notably improved the quality of the final partition. 

For graph-based metrics, the results obtained using the DFN-based partition on 2, 4, 8, and 16 processors were similar to those obtained using the mesh-based partition.
However, the times required for the DFN-based partitions were several orders of magnitude less than the mesh-based partition.
In terms of cut, the quality of the partition projected down from the DFN onto the mesh depends upon the adopted meshing strategy -- uniform resolution or variable resolution. 
In the case of uniform mesh resolution, the projected cuts are along the intersection lines that are the same resolution as the mesh within the fractures. 
However, in the case of a variable resolution mesh, the projection of the DFN partition onto the mesh requires that the cuts be made along the intersections where mesh resolution is finest. 
Due to this, the difference between the cut on $P(G)$ and $P(F)$ is larger than the uniform mesh cases. 
However, this difference can be drastically reduced by refining the partition obtained using the DFN. 
In essence, the additional refinement of the hybrid method provides a good initial condition for the partition solver, and a higher quality partition can be readily obtained.

For computation-based metrics, such as  solver time and FLOPS, the results were similar as well, depending slightly on the adopted meshing scheme. 
The quality of the partition influenced the number of FLOPS, iterations of the Krylov solver, and simulation wall-clock time.   
There was little difference observed in the computational performance of the partitions obtained on the mesh, the DFN, and the refined solution. 
In these test cases, the total computational time was either dominated by the partitioning, in the case of mesh-based partitioning, or the solver, in the case of the DFN based partitioning. 
The difference between the relative contribution of partitioning in the two methods is in stark contrast. 
When combined, the DFN-based partitions resulted in overall run times that were several orders of magnitude lower than the mesh-based partition. 
If we considered these results in terms of ensemble analysis, then the reduction of the overall time for a single moderate-sized DFN realization simulation would allow for an increase in the number of realizations that can be performed at a fixed computational cost. 
In turn, this would lead to the improved bounding of system uncertainty, which is a crucial aspect of subsurface modeling.

In the second test case, we considered a single large DFN containing two and a half thousand fractures meshed with approximately 40 million nodes to test the performance of the proposed method at larger scales.
For this larger network, we partitioned the DFN/mesh into 128, 256, and 512 pieces. 
Here, we determined that the direct partitioning of the mesh based on the partitioning of the DFN was not appropriate. 
The cut and imbalance values were drastically higher than those obtained using the mesh. 
The principal reason for this low quality was that there were not enough fractures to appropriately distribute throughout the partition.
While the lower quality partitions did not affect the runs times significantly in the smaller networks, they did have a significant impact on this larger network. 
In fact, we observed an increase in the simulation run time with an increase in the number of processors due to the increase in communication between processors.  
However, the hybrid method where the DFN-based partition provides a good initial condition for the partition solver, overcame these limitations as it was not limited to the DFN topology.
The refined partition solution provided cut and imbalance values that were close to the mesh-based partition but in a much less time. 
In turn, this method outperformed both of the other methods in terms of the total run time. 
Thus, this additional refinement should be applied as its additional cost is negligible compared to other aspects of the simulation, but the improved performance of the solver is substantial.

The proposed methodology for hijacking multilevel graph partitioning could, in principle, be applied to any system that exhibits a hierarchical multilevel structure and is amenable to a coarse-scale graph representation. 
From a numerical point of view, the key is that the matrix of discretized equations exhibits a strong block structure with few off-diagonal terms.
If this is the case, then a coarse-scale graph representation is a good proxy for the coarsened version of the computational mesh. 
Algebraic multi-grid methods for linear systems also exploit this block structure, and combining the proposed mesh partitioning method with an algebraic multi-grid solver could prove to be a powerful tool. 
This combination will be explored in future studies.
It must be noted that the method was tested on steady single-phase flow solutions. 
In the case of transient flow and transport simulations, the higher cuts will result in more communication between processors as information will be transfered at every time step. 
Therefore, while the direct projection of the DFN-based partition performed similarly to the mesh-based partition in our smaller problem, the additional communication could become problematic in more complex physical systems that are non-linear or time-dependent, or larger systems as we also observed.
Thus, it is essential to apply the additional refinement step in such scenarios due to its marked improvement across all considered metrics with little extra cost.
Furthermore, additional considerations should be taken for flow and particle tracking simulations where the load balancing will need to be dynamic. 
It is possible that the proposed algorithm could be combined with other existing methods, such as ~\citep{walshaw2000multiphase}, to take advantage of the coarse-scale graph representation of the network. 

\section*{Acknowledgments} 
This work was funded by the Department of Energy at Los Alamos National Laboratory through the Laboratory-Directed Research and Development Program LANL LDRD grant \#20170103DR. J.D.H. and M.R.S. acknowledges support from the LANL LDRD program office Grant Number  \#20180621ECR.
Los Alamos National Laboratory is operated by Triad National Security, L.L.C., for the National Nuclear Security Administration of the U.S. Department of Energy (Contract No. 89233218CNA000001).  


\bibliographystyle{spbasic}      
\bibliography{dfn-graph}

\section*{Figures}

\end{document}